\documentclass[useAMS,usenatbib]{mn2e}

\usepackage{graphicx,natbib, bm, amssymb, longtable}

\title[Does the accreting millisecond pulsar XTE J1814$-$338 precess?]{Does the accreting millisecond pulsar XTE J1814$-$338 precess?}
\author[C. T. Y. Chung, D. K. Galloway, A. Melatos]{C. T. Y. Chung$^{1}$, D. K. Galloway$^{1, 2}$, A. Melatos$^{1}$\\
$^{1}$School of Physics, University of Melbourne, Parkville, VIC 3010, Australia\\
$^{2}$School of Physics, Monash University, Clayton, VIC 3800, Australia}

\begin{document}

\date{Accepted 2008 August 22.  Received 2008 August 21; in original form 2008 April 5}

\pagerange{\pageref{firstpage}--\pageref{lastpage}} \pubyear{2007}

\maketitle

\label{firstpage}

\begin{abstract}
Precession in an accretion-powered pulsar is expected to produce characteristic variations in the pulse properties. Assuming surface intensity maps with one and two hotspots, we compute theoretically the periodic modulation of the mean flux, pulse-phase residuals and fractional amplitudes of the first and second harmonic of the pulse profiles. These quantities are characterised in terms of their relative precession phase offsets. We then search for these signatures in 37 days of X-ray timing data from the accreting millisecond pulsar XTE J1814$-$338. We analyse a 12.2-d modulation observed previously and show that it is consistent with a freely precessing neutron star only if the inclination angle is $< 0.1^\circ$, an a priori unlikely orientation. We conclude that if the observed flux variations are due to precession, our model incompletely describes the relative precession phase offsets (e.g. the surface intensity map is over-simplified). We are still able to place an upper limit on $\epsilon$ of $3.0 \times 10^{-9}$ independently of our model, and estimate the phase-independent tilt angle $\theta$ to lie roughly between $5^\circ$ and $10^\circ$. On the other hand, if the observed flux variations are not due to precession, the detected signal serves as a firm upper limit for any underlying precession signal. We then place an upper limit on the product $\epsilon \cos \theta$ of $\leq 9.9 \times 10^{-10}$. The first scenario translates into a maximum gravitational wave strain of $10^{-27}$ from XTE J1814$-$338 (assuming a distance of 8 kpc), and a corresponding signal-to-noise ratio of $\leq 10^{-3}$ (for a 120 day integration time) for the advanced LIGO ground-based gravitational wave detector.
\end{abstract}


\section{Introduction}
\label{intro}
Accreting millisecond pulsars (AMSPs) are a subset of neutron stars in low-mass X-ray binaries (LMXBs) that exhibit persistent X-ray pulsations with periods below 10 ms. In the standard recycling scenario, AMSPs are the evolutionary link between LMXBs and nonaccreting, radio millisecond pulsars \citep{alpar82, radha82}. Eight AMSPs have been discovered at the time of writing \citep{wij04, morgan05, galloway07, krimm07}.

Most AMSPs are X-ray transients. Once every few years, they emerge from quiescence and become detectable during an outburst lasting several weeks. The outburst is attributed to enhanced accretion [e.g. \citet{lasota01}], funnelled onto a small number of hotspots on the star. Little is known about the shape, position, or number of these hotspots \citep{roma04, kulk05}, but they do give rise to detectable X-ray pulsations, from which the spin period and orbital parameters can be determined. During an outburst, surface thermonuclear burning also causes type I X-ray bursts, which last a few minutes and occur on average once every few days. Type I X-ray bursts have been observed in three AMSPs to date: SAX J1808.4$-$3658, XTE J1814$-$338 \citep{wijnands05}, and HETE J1900.1$-$2455 \citep{vanderspek05}. In the burst tails, a small component of the X-ray flux ($\sim 15$\% for XTE J1814$-$338) oscillates at the spin frequency. 

AMSPs are expected to be relatively powerful gravitational wave sources \citep{wattskrishnan08}. The fastest, IGR J00291+5934 \citep{eckert04}, spins at $\Omega_\ast/2\pi = 599$ Hz, well below the theoretical breakup frequency for most nuclear equations of state ($\sim$ 1.5 kHz) \citep{cook94, bildsten98}. Similarly, the fastest radio millisecond pulsar, PSR J1748$-$2446ad \citep{hessels06}, and the fastest nonpulsating LMXB, 4U 1608$-$52 \citep{hartman03}, spin at frequencies of 716 Hz and 619 Hz respectively. The gap below the breakup frequency is explained if the star is deformed by one part in $\sim 10^{8}$, such that gravitational radiation balances the accretion torque at hectohertz frequencies \citep{bildsten98}. Several physical mechanisms can produce the requisite deformation: magnetic mountains \citep{pamel04, melpa05, pamel06, vigmel08}, thermocompositional mountains caused by electron capture gradients \citep{usho00}, toroidal internal magnetic fields \citep{cutler02}, and r-modes \citep{anders98, owen98, nayyar06}. AMSPs are therefore promising targets for ground-based, long-baseline interferometers like the Laser Interferometer Gravitational-Wave Observatory (LIGO). An AMSP at a distance of 1 kpc, spinning at 0.4 kHz with ellipticity $\epsilon = 10^{-8}$, generates a wave strain $h \sim 10^{-27}$. By comparison, initial LIGO's sensitivity threshold in the 0.1--0.4 kHz band is $\sim 10^{-26}$ during the S4 run \citep{abbott07}. Advanced LIGO will get down to $h \sim 10^{-27}$ in the same band, and narrowband tunability will increase its sensitivity to AMSPs further, as $\Omega_\ast$ is known a priori from X-ray timing. 

An AMSP with ellipticity $\epsilon \sim 10^{-8}$ is expected to precess with a period of hours to days.  Magnetic mountains, for example, are built around the magnetic axis, which is misaligned in general with the rotation axis in objects which pulsate \citep{pamel06}. More generally, a mass quadrupole of any provenance should be kicked out of alignment continuously by stochastic accretion torques \citep{joneand02}. Hence AMSPs are promising observational candidates for observing short-period precession. Until now, however, precession has been difficult to detect in neutron stars. Only one source, the radio pulsar PSR B1828$-$11, precesses unambiguously, with period $P_{\rm{p}} =$ 250 d \citep{stalyshem00}. Oscillatory trends in pulse arrival times, with periods of several days, have also been reported tentatively in a few other objects \citep{melatos00, hobbs06, pamel06}, but the physical cause is unclear. 

Free precession consists of a fast wobble about the angular momentum vector $\mathbf{J}$, at approximately the pulsar spin period $P_\ast = 2\pi/\Omega_{\ast}$ and a slow retrograde rotation about the symmetry axis, with period $P_{\rm{p}} = 2 \pi/\Omega_{\rm{p}}$, which modulates the pulse shape and arrival times \citep{zimsz79, alpines85, joneand01, joneand02, link03}. The precession frequency $\Omega_{\rm{p}}$ depends on the ellipticity, $\epsilon$, and the tilt angle $\theta$ (between the symmetry axis and \textbf{J}), with
\begin{equation}
\label{gw}
\epsilon \cos \theta \approx \Omega_p/\Omega_{\star}.
\end{equation}

The amplitude ratio of the gravitational wave signal at the spin frequency and its second harmonic \citep{zimsz79, jara98}, and in the + and $\times$ polarizations, provides \textit{independent} information on $\epsilon, \theta$, the orientation of \textbf{J}, and the emission pattern on the surface of the star. Narrowband tunability facilitates extraction of this information. 

In this paper, we compute theoretically the X-ray signal from a precessing pulsar for a range of orientations and compare three quantities from each pulse profile to the data: the mean flux of the profile, the zero-to-peak pulse amplitude, and the pulse-phase residuals.  We search for the signature of precession in X-ray timing data from one particular AMSP, XTE J1814$-$338. 
An analogous search was carried out by \citet{akgun06} for the radio pulsar PSR B1828-11, who modelled the period residuals and pulse shapes analytically taking into account precession effects (biaxial and triaxial) as well as the contribution from the magnetic spin-down torque. The authors performed searches over a range of beam locations, degrees of triaxiality, tilt angles and angle-dependent spin-down torques, finding a wide range of parameters which match the data. Thus, they were unable to constrain the shape of the star but did find that the angle-dependent spin-down torque contributes to the period residuals. Their method differs from ours in that, instead of fitting the shape of the residuals and comparing for each set of parameters, they determined the validity of a configuration by calculating Bayesian probability distribution functions for the parameters under certain constraints.

 The paper is structured as follows. Section \ref{model} describes the precession model and its implementation. Sections \ref{single} and \ref{double} characterize the predicted X-ray signal for a biaxial, precessing pulsar with one and two hotspots respectively, specifically the relative precession phases between the flux, pulse amplitude, and pulse-phase. Section \ref{triaxial} repeats the predictions for a triaxial, precessing pulsar. Section \ref{datareduction} describes the data reduction and timing analysis of XTE J1814$-$338. We compare the measurements with the theory in Section \ref{compare} and derive upper limits on $\epsilon$, $\theta$, and the associated gravitational wave strain in Section \ref{conclusion}. The limit on $\theta$ constrains the relative strength of the driving and damping forces in the system.

\section{Precession model}
\label{model}
\subsection{Equations of motion}
\begin{figure}
\begin{center}
\includegraphics[scale=0.3]{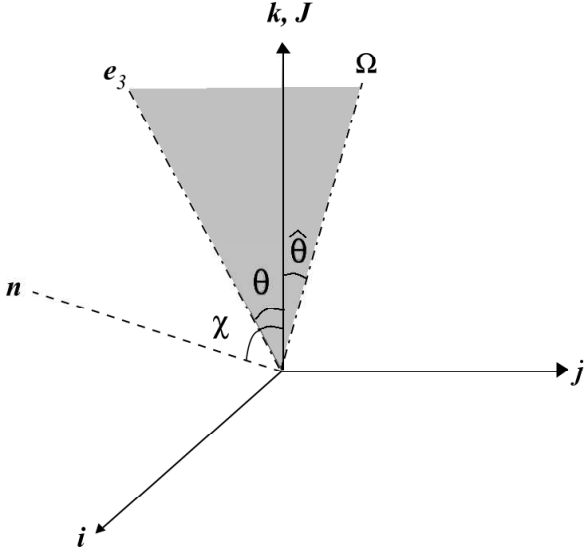}
\caption{Geometry of precession. Drawn are the inertial frame axes {${\mathbf{i}}, {\mathbf{j}}, {\mathbf{k}}$}; the symmetry axis, ${\mathbf{e}}_3$, which makes an angle $\theta$ with the total angular momentum vector, $\mathbf{J}$; the angular velocity vector, $\mathbf{\Omega}$, which makes an angle $\hat{\theta}$ with \textbf{J}; and the line-of-sight vector \textbf{n}, which makes an angle $\chi$ with $\mathbf{J}$. The dashed line indicates that \textbf{n} lies in the plane containing $\bm{i}$ and $\bm{k}$. The dashed-dotted lines indicate that ${\mathbf{e}}_3$ and $\Omega$ are coplanar. The shaded plane containing $\mathbf{e}_3, \bm{\Omega}$, and \textbf{J} rotates about \textbf{J} with angular frequency $\dot{\phi}$. }
\label{figure1}
\end{center}
\end{figure}

Three Euler angles ($\theta, \phi, \psi$) describe the rotation of a rigid body with body axes ($\mathbf{e}_1, \mathbf{e}_2, \mathbf{e}_3$) relative to the Cartesian triad ($\mathbf{i}, \mathbf{j}, \mathbf{k}$) of an inertial observer. We define ($\theta, \phi, \psi$) according to the \citet{lanlif69} convention. Consider, first, the special case of a biaxial, freely precessing neutron star. Let ${\mathbf{e}}_3$ be the symmetry axis and take ${\mathbf{k}}$ to lie along the total angular momentum $\mathbf{J}$, as depicted in Figure \ref{figure1}. The angle between $\mathbf{e}_3$ and \textbf{J} is $\theta$. The total angular velocity of the precessing system, $\mathbf{\Omega}$, comprises two components: $\mathbf{\Omega}$ rotates about $\mathbf{J}$ at a constant angle ${\hat{\theta}}$, with frequency $\dot{\phi}$, and ${\mathbf{e}}_1$ and ${\mathbf{e}}_2$ rotate about $\mathbf{e}_3$, with frequency $\Omega_{p} = \dot{\psi}$. The rate of precession is controlled by the ellipticity of the star, $\epsilon$, and $\theta$, i.e. $\Omega_{p} = \epsilon \Omega \cos (\theta + \hat{\theta})$. Note that $\mathbf{J}, \mathbf{\Omega}$, and $\mathbf{e}_3$ are coplanar, as indicated by the shading in Figure \ref{figure1}. For small angles, one has $\hat{\theta} \approx (\Delta I_{\rm{d}}/I_1) \sin \theta \cos \theta$, where $I_1, I_2$ and $I_3$ denote the star's principal moments of inertia, and $\Delta I_{\rm{d}}$ is defined through $I_1 = I_2 = I_0 - \Delta I_{\rm{d}}/3$ and $I_3 = I_0 + 2 \Delta I_{\rm{d}}/3$. $\Delta I_{\rm{d}}$ is positive for an oblate star, negative for a prolate star, and is related to the ellipticity via $\epsilon = \Delta I_{\rm{d}}/I_0$.

In terms of the above notation, the six equations of motion are
\begin{eqnarray}
\label{thetadot}
\dot{\theta} &=& \Omega_1 \cos\psi  - \Omega_2 \sin\psi, \\
\dot{\phi} &=& (\Omega_1 \sin\psi + \Omega_2 \cos\psi )/\sin\theta,\\
\dot{\psi} &=& \Omega_3 - {\cos\theta} (\Omega_1 \sin\psi + \Omega_2 \cos\psi)/\sin\theta, \\\label{o1}
I_1 \dot{\Omega_1} &=& (I_2 - I_3) \Omega_2 \Omega_3, \\ 
I_2 \dot{\Omega_2} &=& (I_3 - I_1) \Omega_1 \Omega_3, \\ \label{o2}
I_3 \dot{\Omega_3} &=& (I_1 - I_2) \Omega_1 \Omega_2. \label{o3}
\end{eqnarray}
The time origin is arbitrary, so initially we can set $\psi(0) = 0$ without loss of generality. The angular velocity can be decomposed into $\bm{\Omega} = \dot{\phi}\bm{k} + \dot{\psi}\bm{e}_3$. This gives $\dot{\psi} = -\epsilon \dot{\phi} I_0/I_3$. In biaxial precession, $\theta$ is constant. Solving (\ref{thetadot})--(\ref{o3}) with $\dot{\theta} = 0$ and $\psi(0) = 0$ yields the following expressions, which we use to initialise the angular velocity:

\begin{eqnarray}
\Omega_{1}(0) &=& 0,\\
\Omega_{2}(0) &=& \Omega_c \sin\theta, \\
\Omega_{3}(0) &=& \Omega_c (\cos\theta - \epsilon I_0/I_3).
\end{eqnarray}

Without loss of generality, we fix the line-of-sight vector, \textbf{n}, to lie in the \textbf{i}-\textbf{k} plane, making an angle $\chi$ with $\mathbf{J}$ and intersecting the stellar surface intensity map at latitude $\theta_{\rm{B}} $ and phase $\phi_{\rm{B}} $. The angles ($\theta_{\rm{B}}$,$\phi_{\rm{B}}$) are defined with respect to the moving body frame, as opposed to the Euler angles, which are defined with respect to the inertial frame. We then compute the observed intensity $I$ as a function of time from a specific surface intensity map $F(\theta_{\rm{B}}, \phi_{\rm{B}})$, i.e. $I = F(\theta_{\rm{B}}, \phi_{\rm{B}})$. Note that \citet{joneand01} assumed $0 \leq \chi \leq \pi/2$ but did not investigate the dependence of $I$ on $\chi$. In this paper, we show that the $\chi$ dependence is significant.

The above initialization gives four searchable parameters: $\theta$, the initial azimuth $\phi(0)$, the inclination angle $\chi$, and the latitude $\alpha$ of the hotspot(s), defined in Section \ref{profiles}.

\subsection{Synthetic pulse profiles}\label{profiles}
\begin{figure}
\begin{center}
\includegraphics[scale=0.5]{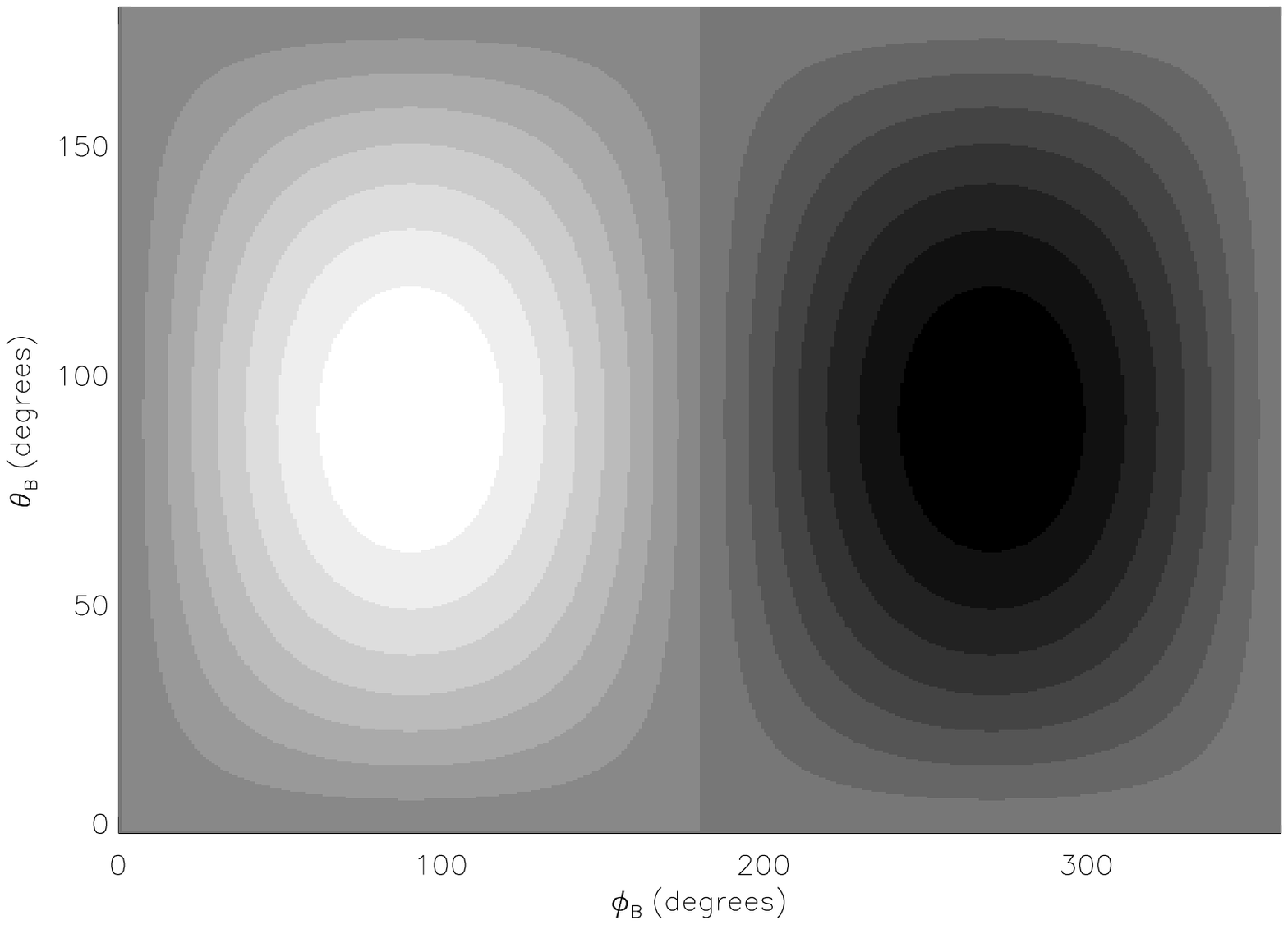}
\hfill
\includegraphics[scale=0.5]{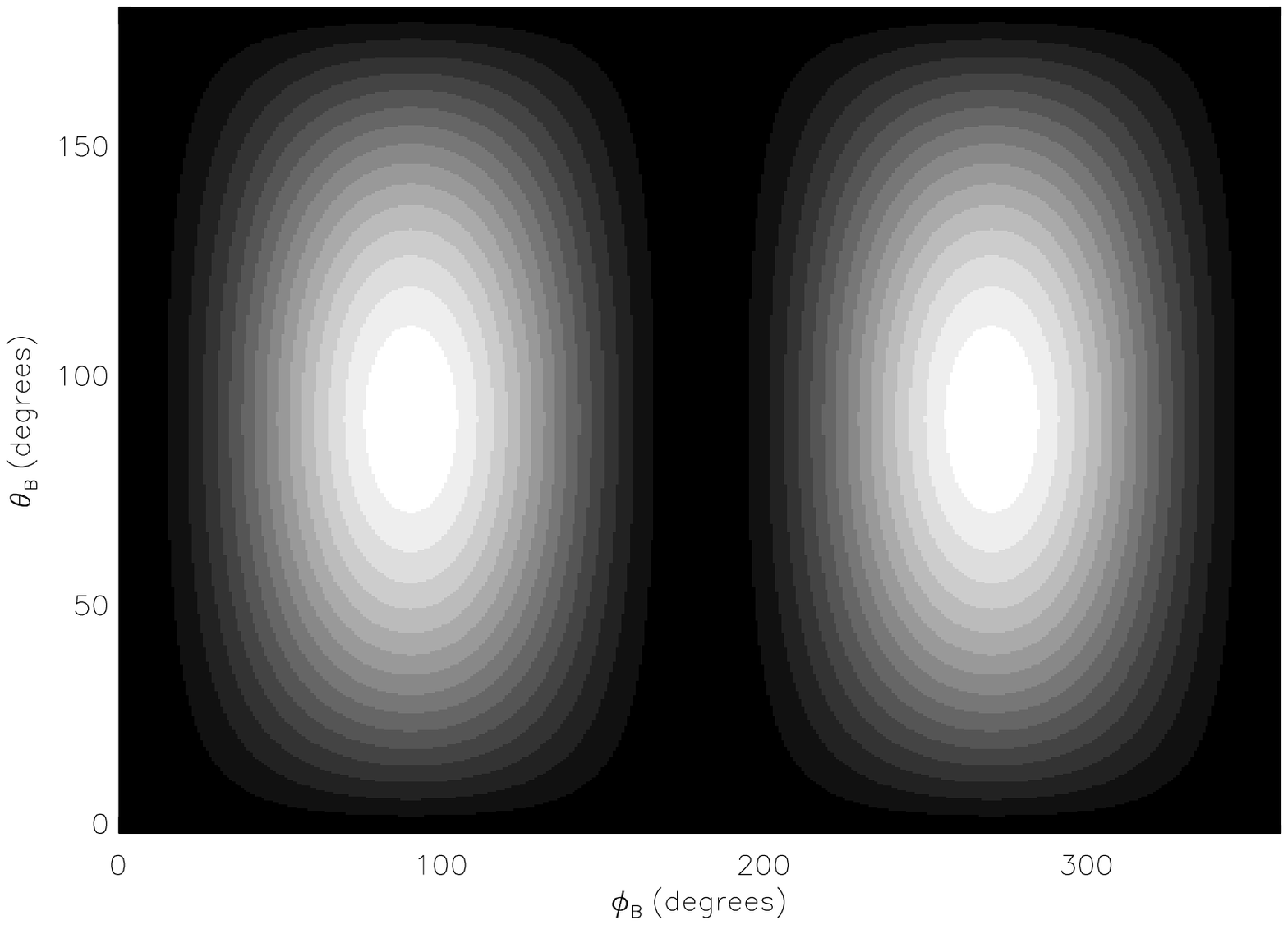}
\caption{Top: Surface intensity map of the single hotspot configuration (equation \ref{hotspot1}). Bottom: Surface intensity map of the double hotspot configuration (equation \ref{hotspot2}).}
\label{figure2}
\end{center}
\end{figure}

Studies of the harmonic ratio and modulation amplutude of type I burst oscillations from six LMXBs point to the existence of a single, hemispheric hotspot \citep{muno02, pamel06}.
We thus perform simulations for one and two hotspots, corresponding to 
\begin{equation}
\label{hotspot1}
F(\theta_{\rm{B}}, \phi_{\rm{B}}) = \sin(\theta_{\rm{B}} + \alpha) \sin(\phi_{\rm{B}}) + \rm{DC}
\end{equation}
and
\begin{equation}
\label{hotspot2}
F(\theta_{\rm{B}}, \phi_{\rm{B}}) = \sin(\theta_{\rm{B}} + \alpha)\sin^2(\phi_{\rm{B}}) + \rm{DC}
\end{equation}
respectively, where DC represents a constant offset, and $\alpha$ is the latitude of the hotspot's centre, defined to be zero at the equator and $\pm 90^\circ$ at the poles. For $\alpha = 0$, this corresponds to a surface intensity map containing a bright spot on the $\mathbf{e}_1$-$\mathbf{e}_3$ plane, centred on the equator at body coordinates $(\theta_B, \phi_B)=$($90^\circ, 90^\circ$), and a dark or bright spot diametrically opposite at body coordinates ($90^\circ, 270^\circ$) as per Figure \ref{figure2}. For $\alpha \ne 0$, the body coordinates in (\ref{hotspot1}) and (\ref{hotspot2}) are rotated by $\alpha$ in the $\mathbf{e}_1$-$\mathbf{e}_3$ plane.

The light curves are generated by evaluating the intensity of the point on the surface map which is intercepted by the line-of-sight at each time step. These synthesised light curves are analysed in exactly the same way as the observational data. 
The profile of each pulse in the time series is fitted with a first and second harmonic, i.e.  $a + b \sin (2 \pi \gamma + c) + d \sin (4 \pi \gamma + e)$, where $a$ is the mean flux, $\gamma$ denotes the pulse phase ($0 \leq \gamma \leq 1$), $b$ and $d$ are amplitudes, and $c$ and $e$ are phase offsets. The fractional root-mean-square (RMS) amplitudes of the first and second harmonic are $b/\sqrt{2}a$ and $d/\sqrt{2}a$ respectively. We define two reference pulse phases in each profile as the phases which maximise the first and second harmonic components. The pulse-phase residuals are the differences between the predicted and observed maxima, i.e. $\gamma_0 = 1/4 - c/2 \pi$ and $\gamma_1 = 1/8 - e/4\pi$. In Section \ref{firstharm}, we show that the second harmonic does not contribute to the overall candidate precession signal at a level that can be detected in the X-ray timing data presently available for XTE J1814$-$338. The quantities $d/\sqrt{2}a$ and $\gamma_1$ are therefore neglected in the following analysis. 

\begin{figure}
\begin{center}
\includegraphics[scale=0.4]{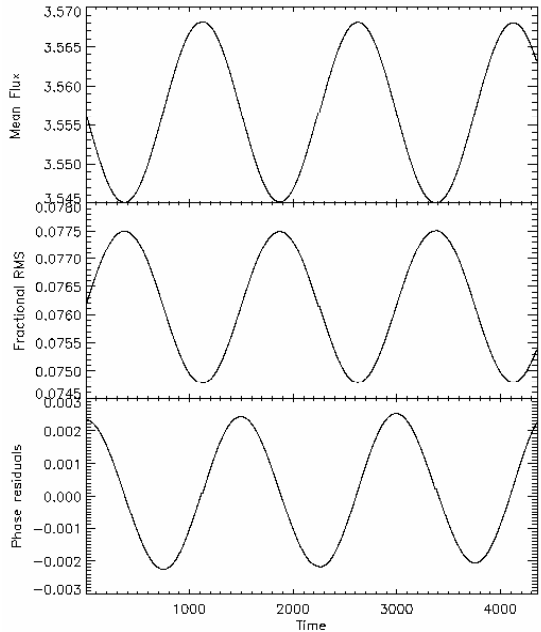}
\caption{Sample simulation output for a biaxial star with a single hotspot. Top: mean flux, $a$. Center: fractional RMS, $b/\sqrt{2}a$. Bottom: pulse-phase residuals, $\gamma_0$. The precession period is 1500 time units.}
\label{figure3}
\end{center}
\end{figure}

We fold the time series $a$, $b/\sqrt{2}a$ and $\gamma_0$ at the theoretical precession period. These three quantities are modulated at the precession frequency due to the motion of the hotspot(s) relative to the observer (see Figure \ref{figure3} for an example of the output). The average trend in each quantity is fitted with a sinusoid, viz. $A_n + B_n \sin (2 \pi \Gamma + C_n)$, where $\Gamma$ is the precession phase, not to be confused with the pulse phase, and $n$ refers to mean flux, fractional RMS or pulse-phase residuals. This yields three main quantities of observational interest: the relative precession phase of $a$ and $b/\sqrt{2}a$ (denoted by $\Delta \Gamma_{\rm{flux-rms}} = |C_{\rm{flux}} - C_{\rm{rms}}|$), the relative precession phase $\gamma_0$ and $b/\sqrt{2}a$ (denoted by $\Delta \Gamma_{\rm{phase-rms}} = |C_{\rm{phase}} - C_{\rm{rms}}|$), and the zero-to-peak amplitude of the pulse-phase residuals ($B_{\rm{phase}}$). Our simulations run for (on average) three precession periods.

\section{Biaxial star: Single hotspot}
\label{single}

In most configurations involving a single hotspot, the precession phases of $b/\sqrt{2}a$ and $\gamma_0$ differ by $\Delta \Gamma_{\rm{phase-rms}} \approx \pi/2$. The precession phases of $b/\sqrt{2}a$  and $a$ are either in phase or antiphase, i.e. $\Delta \Gamma_{\rm{flux-rms}} = 0$ or $\pi$. $B_{\rm{phase}}$ increases with the tilt angle $\theta$. 

The effect of varying the four searchable parameters is now discussed in detail. 

\subsection{Tilt angle, $\theta$} 
We consider small tilt angles ($\theta \leq 10^\circ$), as in PSR 1828$-$11 [$\theta < 3^\circ$; \citet{stalyshem00, link03}] and other astrophysical bodies.
We verify that the precession period increases with $\theta$ according to (\ref{gw}). We also find that $B_{\rm{phase}}$ increases linearly with $\theta$ in the small tilt angle regime $\theta \leq 10^\circ$, ranging from 0.003 at $\theta = 1^\circ$ to 0.028 at $\theta = 10^\circ$. There is also a dependence of $B_{\rm{phase}}$ on $\alpha$ (see bottom panel of Figure \ref{figure5}). 
\citet{joneand01} predicted analytically, for a radio pulsar with a directed beam rather than a hotspot, that the pulse-phase residuals vary sinusoidally on the precession time-scale with amplitude $B_{\rm{phase}} \propto \theta/\tan (\pi/2 - \alpha')$, where $\alpha'$ in their case is the latitude of the beam. This formula agrees with our results.

\begin{figure}
\includegraphics[scale=0.85]{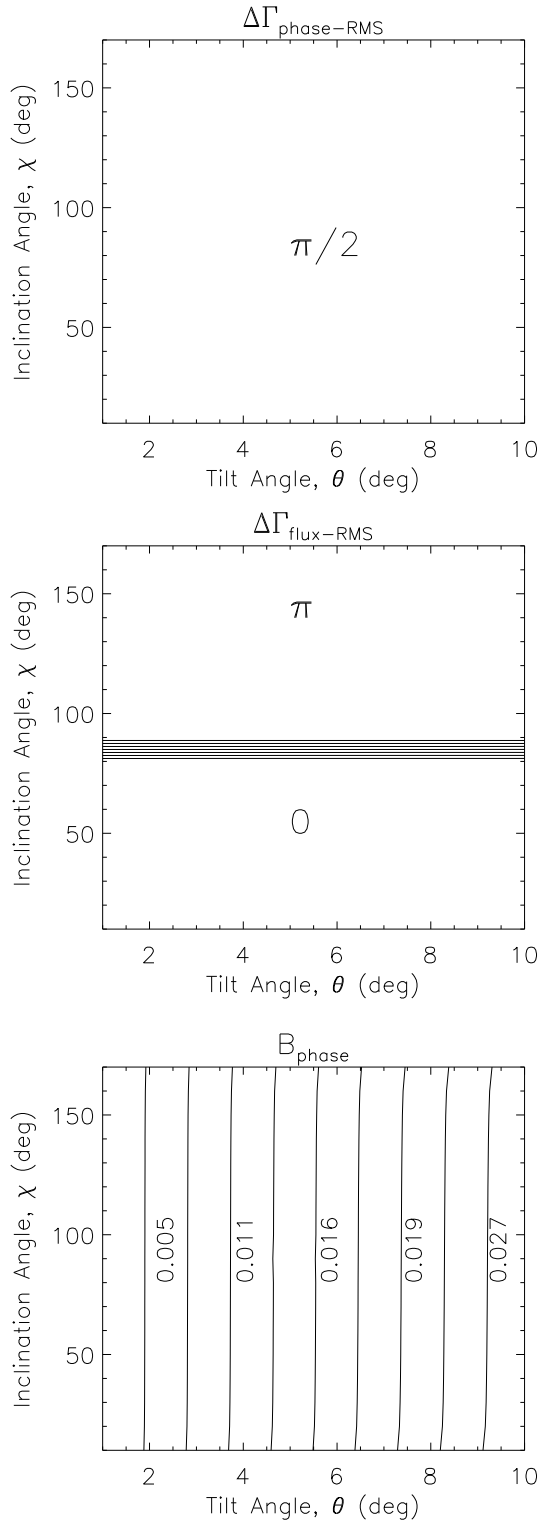}
\caption{Relative precession phase of the RMS and pulse-phase residuals, $\Delta \Gamma_{\rm{phase-rms}}$ (top), RMS and mean flux, $\Delta \Gamma_{\rm{flux-rms}}$ (middle), and amplitude of the folded pulse-phase residuals, $B_{\rm{phase}}$ (bottom), versus tilt angle $\theta$ and inclination angle $\chi$, both measured in degrees for a biaxial pulsar with one hotspot. Parameters: $\phi_{(0)} = 0^\circ$, $\alpha = 45^\circ$ for $1^\circ \leq \theta \leq 10^\circ$, and $5^\circ \leq \chi \leq 175^\circ$.}
\label{figure4}
\end{figure}

 Varying $\theta$ does not affect $\Delta \Gamma_{\rm{flux-rms}}$ or $\Delta \Gamma_{\rm{phase-rms}}$. $\Delta \Gamma_{\rm{flux-rms}}$ undergoes a phase shift of $\pi$ around $\chi = 90^\circ$ which is explained below. Contour maps of $\Delta \Gamma_{\rm{flux-rms}}$, $\Delta \Gamma_{\rm{phase-rms}}$ and $B_{\rm{phase}}$ as functions of $\chi$ and $\theta$ are plotted in Figure \ref{figure4}. 

\subsection{Hotspot latitude, $\alpha$, and inclination, $\chi$}\label{alphachi1}
\begin{figure}
\includegraphics[scale=0.85]{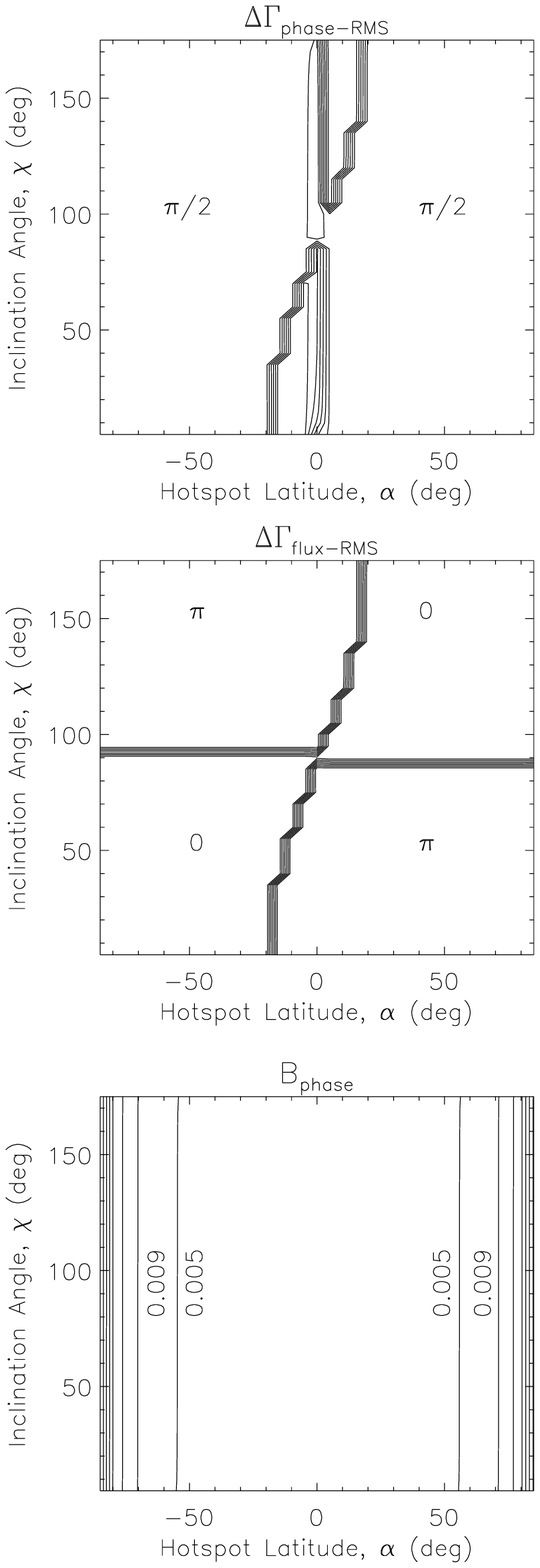}
\caption{Relative precession phase of the RMS and pulse-phase residuals, $\Delta \Gamma_{\rm{phase-rms}}$ (top), RMS and mean flux, $\Delta \Gamma_{\rm{flux-rms}}$ (middle), and amplitude of the folded pulse-phase residuals, $B_{\rm{phase}}$ (bottom), versus inclination angle $\chi$ and hotspot latitude $\alpha$,  both measured in degrees for a biaxial pulsar with one hotspot. Parameters: $\theta = 1^\circ$, $\phi_{(0)} = 0$, $5^\circ \leq \alpha \leq 175^\circ$, and $5^\circ \leq \chi \leq 175^\circ$.}
\label{figure5}
\end{figure}

Figure \ref{figure5} displays contour maps of $\Delta \Gamma_{\rm{phase-rms}}, \Delta \Gamma_{\rm{flux-rms}}$, and $B_{\rm{phase}}$ as function of $\chi$ and $\alpha$ for $\theta = 1^\circ$. In the range $-15^\circ < \alpha < 15^\circ$, $\Gamma_{\rm{RMS}}$ undergoes a $\pi$ phase reversal between certain values of $\chi$, which shows up in $\Delta \Gamma_{\rm{phase-rms}}$ and $\Delta \Gamma_{\rm{flux-rms}}$.
As $\chi$ is increased, a strong second harmonic gradually appears in $\Gamma_{\rm{RMS}}$ over an interval of $\approx 10^\circ$. As $\chi$ increases further, the harmonic disappears and $\Gamma_{\rm{RMS}}$ is shifts by $\pi$. The value of $\chi$ at which this happens depends on $\alpha$ (see top and middle panels of Figure \ref{figure5}). For example, at $\alpha = -10^\circ$, the harmonic appears for $56^\circ < \chi < 64^\circ$. 

The value of $\Delta \Gamma_{\rm{flux-rms}}$ also depends on which hemisphere (north or south) the hotspot is in, relative to \textbf{n}. If both are in the same hemisphere, then $\Delta \Gamma_{\rm{flux-rms}} \approx \pi$. If they are in different hemispheres, then $\Delta \Gamma_{\rm{flux-rms}} \approx 0$. 
This change occurs because the mean flux profile `flips'. For example, if $\chi = 30^\circ$, the fractional RMS is the same whether the hotspot is in the north or south. However, if the hotspot is in the north (south), \textbf{n} starts off closer to (further from) the hotspot, and the flux dims (brightens) as the star precesses. The phase shifts of $\Gamma_{\rm{RMS}}$ and the mean flux are explained geometrically in more detail in Section \ref{stepexplain1}.

$B_{\rm{phase}}$ is larger when the hotspot is close to the poles than when the hotspot is near the equator, but hardly varies with $\chi$. When $\chi = 0$, the observer does not see any evidence of precession at all in the pulse-phase residuals, as $\mathbf{e}_3$ remains equidistant from \textbf{n} at all times.

\subsection{Initial longitude, $\phi(0)$}

$\phi(0)$ determines the initial latitude where \textbf{n} intersects the surface. When \textbf{n} is within $1^\circ$ of $\mathbf{J}$, i.e. $\chi < 1^\circ$ or $\chi > 179^\circ$, as one various $\phi(0)$ from 0 to $2\pi$, $\Delta \Gamma_{\rm{phase-rms}}$ oscillates sinusoidally around $\pi/2$, $\Delta \Gamma_{\rm{flux-rms}}$ oscillates sinusoidally around 0 or $\pi$ (depending on the hemisphere of the hotspot), and $B_{\rm{phase}}$ peaks at $\phi(0) = \pi$ or $2\pi$. At these extremes, \textbf{n} traverses a very small area on the star's surface during each spin period, i.e. $B_{\rm{rms}}$, $B_{\rm{phase}} \ll 1$. Hence any variation due to $\phi(0)$, insignificant for other values of $\chi$, now dominates. 

At $\chi = 90^\circ$, $\Delta \Gamma_{\rm{flux-rms}}$ varies gradually from 0 to $\pi$. This is due to the flip in the mean flux profile mentioned in Section \ref{alphachi1}.

\subsection{Geometry of the phase shifts}\label{stepexplain1}
\begin{figure*}
\includegraphics[scale=0.35]{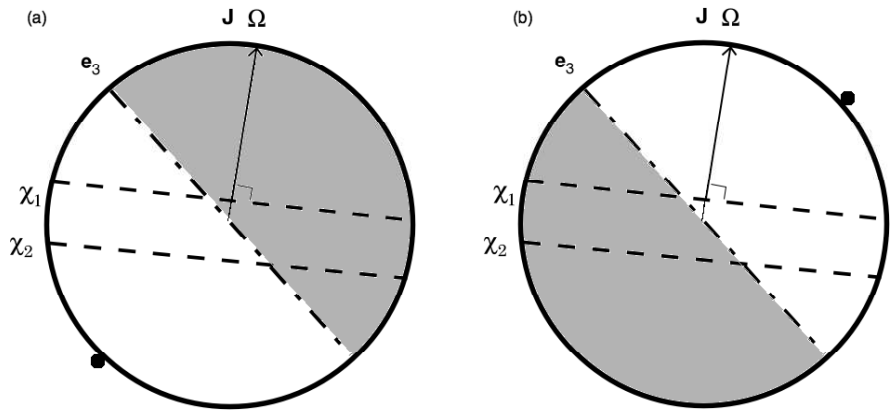}
\caption{Schematic illustrating the rotation of one hotspot, from (a) the start of one precession cycle, to (b) midway through the precession cycle. The bright and dark hemispheres are marked. $\chi_1$ is the position of an observer just north of the equator ($\chi < 90^\circ$). $\chi_2$ is the position of an observer just south of the equator ($\chi > 90^\circ$). The dashed lines indicate the points on the surface that the observers see each spin period. These lines are perpendicular to $\mathbf{\Omega}$ and hence tilted by the angle $\hat{\theta}$ (artificially enhanced for clarity) with respect to the horizontal. The hotspot centres are indicated by the black dots.}
\label{figure6}
\end{figure*}

In this section, we explain geometrically why the relative precession phase of maximum rotation-averaged intensity abruptly changes whenever the line of sight crosses the equator, causing $\Delta \Gamma_{\rm{flux-rms}}$ to jump by $\pi$ rad. 
To understand this counterintuitive effect, consider Figure \ref{figure6}, which shows the coplanar vectors \textbf{J}, \textbf{e$_3$} (which makes an angle $\theta$ with \textbf{J}), and $\mathbf{\Omega}$ (which makes an angle $\hat{\theta}$ with \textbf{J}). In reality, $\hat{\theta}$ is small, but we enlarge it artificially for illustrative purposes. For $\alpha = 0$, the hotspot is at the equator (relative to \textbf{e$_3$}), and the surface intensity changes from brighter to darker than average at the dot-dashed line. Figure \ref{figure6}(a) is a snapshot taken of the pulsar at the start of a precession cycle, while Figure \ref{figure6}(b) is taken half a precession cycle later, after a time $\pi/\dot{\psi}$ elapses. Both snapshots are taken at the same arbitrary pulse-phase, when \textbf{e$_3$} and $\mathbf{\Omega}$ line up as shown. 

Now consider two observers, $\chi_1$ just north of the equator and $\chi_2$ just south of the equator. The dashed lines show the loci of points where the lines-of-sight intersect the surface during one spin (not precession) period. The lines are tilted with respect to the horizontal at angle $\hat{\theta}$, perpendicular to $\mathbf{\Omega}$, because the pulsar rotates instantaneously about $\mathbf{\Omega}$ during one spin period.
At the start of the precession cycle [panel (a)], observer $\chi_1$ traces a path which passes through more of the dark hemisphere than the bright, while $\chi_2$ traces a path through more of the bright hemisphere. Half a precession cycle later [panel (b)], the opposite happens. This means that the mean flux seen by $\chi_1$ increases from minimum to maximum from (a) to (b), whereas the mean flux seen by $\chi_2$ decreases from maximum to minimum. This `flip' causes the $\pi$ phase shift in $\Delta \Gamma_{\rm{flux-rms}}$. 

The phase reversal in $\Gamma_{\rm{RMS}}$ at certain inclination angles when the hotspot is in the range $-15^\circ < \alpha < 15^\circ$ can also be understood with the help of Figure \ref{figure6}. For $\alpha = 0$, the line dividing the bright and dark hemispheres is at the same position in (a) and (b). Hence, the pulse amplitude $b$ and times-of-arrivals (TOAs)  seen by either observer $\chi_1$ or $\chi_2$ is the same at epoch (a) and epoch (b), even though the mean flux $a$ changes from epoch (a) to epoch (b). As we are measuring $b/\sqrt{2}a$, the phase reversal in the mean flux causes a phase reversal in $\Gamma_{\rm{RMS}}$ at $\chi = 90^\circ$. This effect occurs at larger (smaller) $\chi$ for $\alpha > 0^\circ (\alpha < 0^\circ)$, and does not occur when $\alpha > 15^\circ$. This is because, as the hotspot moves away from $\alpha = 0^\circ$, the pulse amplitudes $b$ seen by $\chi_1$ and $\chi_2$ become increasingly different at epochs (a) and (b), so the effect of $a$ on the fractional RMS is reduced. 

We also observe a strong harmonic component in the phase residuals at $\alpha = 0^\circ$ as the pulse TOAs are identical in epochs (a) and (b).

\section{Biaxial star: Two hotspots}
\label{double}

To simulate two diametrically opposed hotspots, the intensity map is changed to equation (\ref{hotspot2}). The pulse profiles now contain an increased harmonic component in certain configurations. The parameter study in Section \ref{single} is repeated.

\subsection{Tilt angle, $\theta$}
Figure \ref{figure7} shows contour maps of the three quantities as a function of $\theta$ and $\chi$ for hotspots at $45^\circ$ and $225^\circ$. As with one hotspot, only $B_{\rm{phase}}$ is affected by changing $\theta$; it ranges from 0.002 to 0.075 (see bottom panel of Figure \ref{figure7}). $d B_{\rm{phase}}/ d \theta$ is smaller in the range $45^\circ \lesssim \chi \lesssim 135^\circ$.

\subsection{Hotspot latitude, $\alpha$ and inclination, $\chi$}
The positions of the hotspots relative to \textbf{n} affect $\Delta \Gamma_{\rm{phase-rms}}$ and $\Delta \Gamma_{\rm{flux-rms}}$. Figure \ref{figure8} shows contour maps of the two quantities and $B_{\rm{phase}}$ as functions of $\chi$ and $\alpha$ for $\theta = 1^\circ$. We find $\Delta \Gamma_{\rm{phase-rms}} \approx \pi/2$ in the regions where $\chi$ is north of a hotspot in the northern hemisphere, or south of one in the southern hemisphere, i.e. if $\alpha = 45^\circ, 225^\circ$ and $\chi < 45^\circ$ or $\chi > 135^\circ$. Elsewhere, we find $\Delta \Gamma_{\rm{phase-rms}} \approx 3\pi/2$. As for one hotspot, the fractional RMS reverses phase when \textbf{n} lies within $\sim 5^\circ$ of the hotspots, causing a second harmonic component to develop in the fractional RMS.

As $\chi$ approaches the equator, the harmonic component in the pulse profiles increases and peaks at $\chi = 90^\circ$. The horizontal band across the contour map at $\chi = 90^\circ$ is caused by the pulse-phase residuals shifting by $\pi$ (see top panel of Figure \ref{figure8}). At $\alpha = 0^\circ$, the two hotspots are within $\theta$ of \textbf{n}. For small $\theta$, the observer sees the fractional RMS, mean flux and pulse-phase residuals rise and fall twice as the star precesses about $\mathbf{e}_3$ (at rate $\dot{\psi}$). This halves the apparent precession period.  

$\Delta \Gamma_{\rm{flux-rms}}$ varies from 0 to $\pi$ in a similar fashion. The fractional RMS and mean flux are generally out of phase when \textbf{n} is north of a hotspot in the northern hemisphere, south of a hotspot in the southern hemisphere, and vice versa. However, the fractional RMS and mean flux undergo phase shifts at different points ($\chi, \alpha$). The fractional RMS `flips' at the points indicated by the diamond pattern in the top panel of Figure \ref{figure8}, whereas the mean flux `flips' for $65^\circ \lesssim \chi \lesssim 115^\circ$, indicated by the horizontal band in the middle panel of Figure \ref{figure8}. The origin of these phase shifts is explained geometrically in Section \ref{explainstep2}.

$B_{\rm{phase}}$ is largest when both $\chi$ and $\alpha$ are close to the poles.

\subsection{Initial longitude, $\phi(0)$}

Longitudinal dependences are weak. For example, $\Delta \Gamma_{\rm{phase-rms}}$ varies by $< 0.02\%$ as $\phi(0)$ goes from $0^\circ$ to $330^\circ$ for $\chi = 70^\circ$, as is shown in Figure \ref{figure9}. 
Near $\chi = 90^\circ$, $\Delta \Gamma_{\rm{phase-rms}}$ and $\Delta \Gamma_{\rm{flux-rms}}$ vary more strongly with $\phi(0)$ due to phase shifts in the pulse-phase residuals and fractional RMS at various configurations. However, these shifts do not translate into true changes in the observed pulse profile, because the second harmonic component, which we do not consider, dominates in this region and behaves differently. The mean flux profile is not affected. $\Delta \Gamma_{\rm{flux-rms}}$ cycles between $\approx 0.9$ and $2\pi$ as $\phi(0)$ varies from 0 to $2\pi$ rad, while $\Delta \Gamma_{\rm{phase-rms}}$ cycles between $\pi/2$ and $3\pi/2$.

\begin{figure}
\includegraphics[scale=0.85]{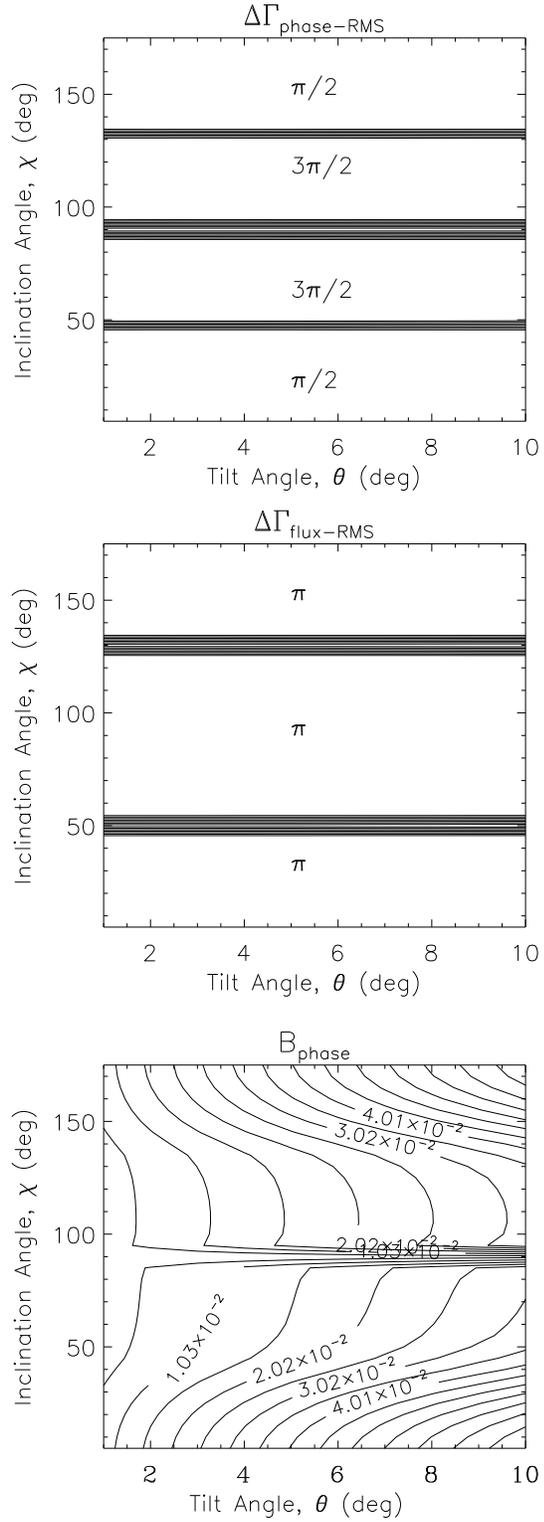}
\caption{Relative precession phase of the RMS and pulse-phase residuals, $\Delta \Gamma_{\rm{phase-rms}}$ (top), RMS and mean flux, $\Delta \Gamma_{\rm{flux-rms}}$ (middle), and amplitude of the folded pulse-phase residuals, $B_{\rm{phase}}$ (bottom), versus inclination angle $\chi$ and tilt angle $\theta$,  both measured in degrees, for a biaxial pulsar with two hotspots. Parameters: $\phi(0) = 0$, $\alpha = 45^\circ$, $1^\circ \leq \theta \leq 10^\circ$ and $5^\circ \leq \chi \leq 175^\circ$.}
\label{figure7}
\end{figure}

\begin{figure}
\includegraphics[scale=0.85]{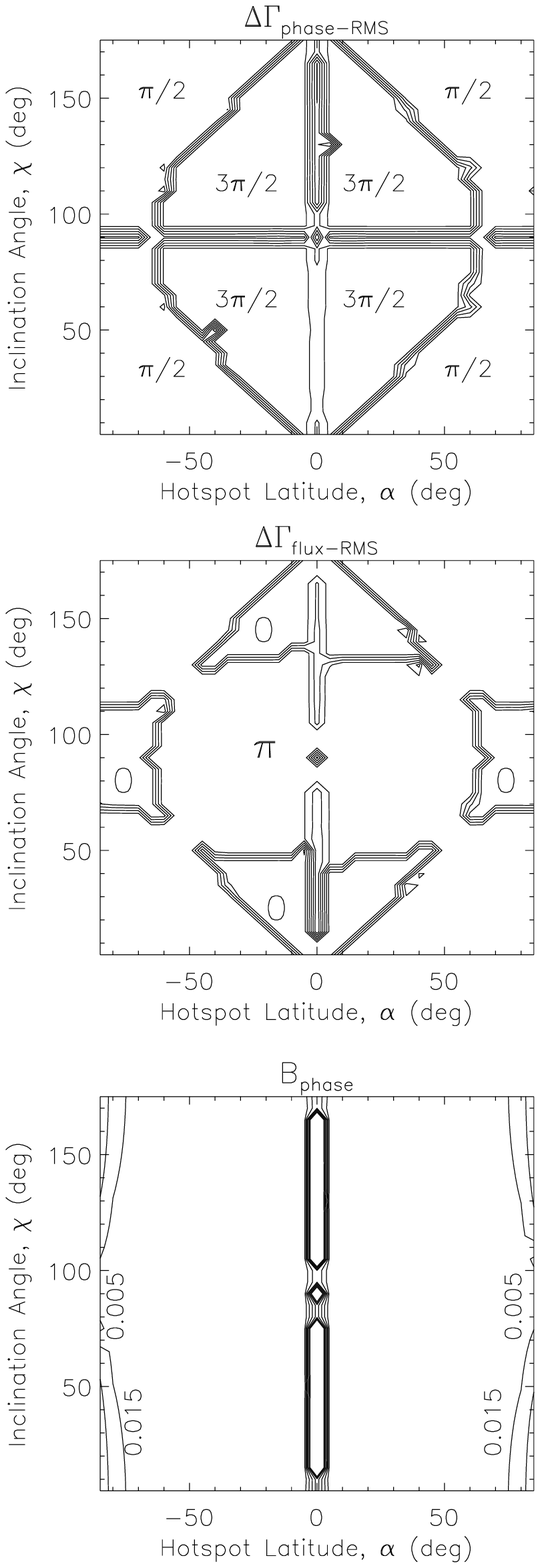}
\caption{Relative precession phase of the RMS and pulse-phase residuals, $\Delta \Gamma_{\rm{phase-rms}}$ (top), RMS and mean flux, $\Delta \Gamma_{\rm{flux-rms}}$ (middle), and amplitude of the folded pulse-phase residuals, $B_{\rm{phase}}$ (bottom), versus inclination angle $\chi$ and hotspot latitude $\alpha$,  both measured in degrees, for a biaxial pulsar with two hotspots. Parameters: $\theta = 1^\circ$, $\phi(0) = 0$, $5^\circ \leq \alpha \leq 175^\circ$ and $5^\circ \leq \chi \leq 175^\circ$.}
\label{figure8}
\end{figure}

\begin{figure}
\includegraphics[scale=0.85]{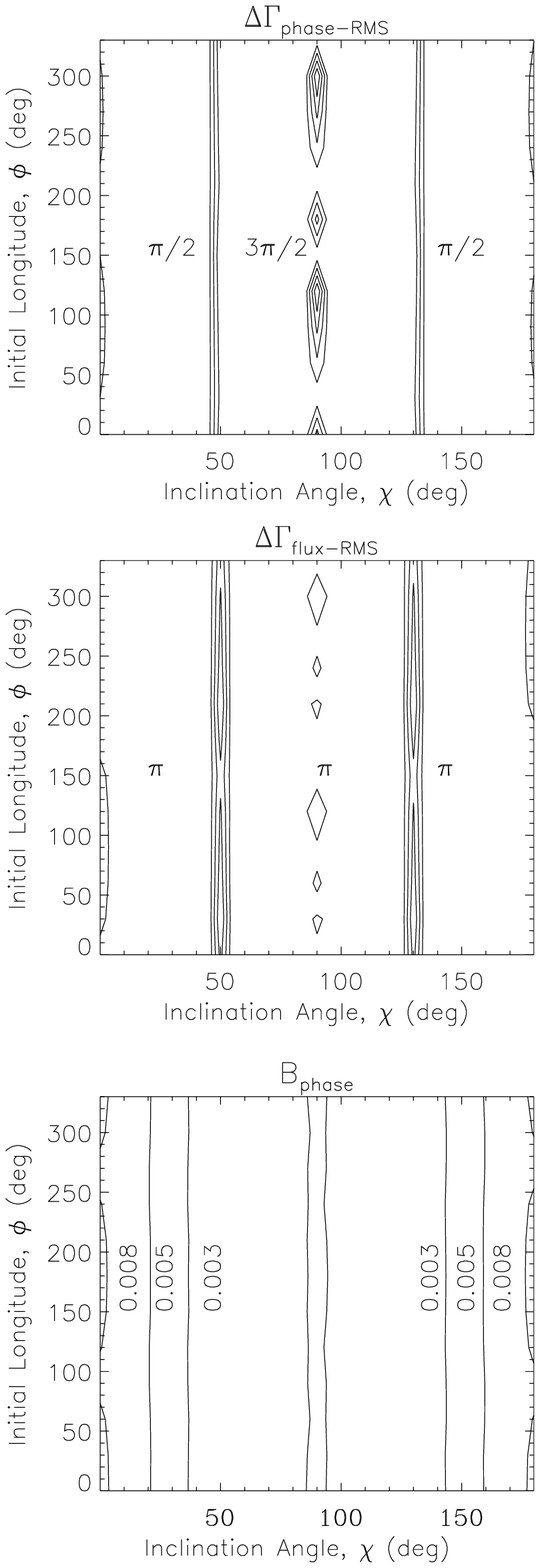}
\caption{Relative precession phase of the RMS and pulse-phase residuals, $\Delta \Gamma_{\rm{phase-rms}}$ (top), RMS and mean flux, $\Delta \Gamma_{\rm{flux-rms}}$ (middle), and amplitude of the folded pulse-phase residuals, $B_{\rm{phase}}$ (bottom), versus inclination angle $\chi$ and initial phase $\phi(0)$,  both measured in degrees, for a biaxial pulsar with two hotspots. Parameters: $\theta = 1^\circ$, $\alpha = 45^\circ$, $0^\circ \leq \phi \leq 330^\circ$ and $0.1^\circ \leq \chi \leq 179.9^\circ$.}
\label{figure9}
\end{figure}

\subsection{Geometry of the phase shifts}\label{explainstep2}
\begin{figure*}
\includegraphics[scale=0.35]{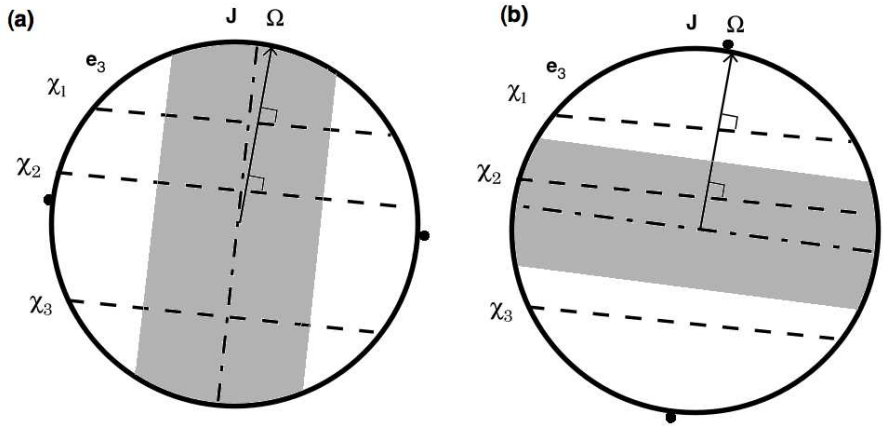}
\caption{Schematic illustrating the rotation of two hotspots, from (a) the start of one precession cycle, to (b) midway through the precession cycle. The shaded band indicates the region on the surface that is darger than average, for $\alpha \approx 45^\circ$. $\chi_1$ is the position of an observer just north of a hotspot ($\chi < 45^\circ$), $\chi_2$ is the position of an observer south of a hotspot ($\chi > 65^\circ$). $\chi_3$ is the position of an observer south of $115^\circ$, where the mean flux profile is seen to flip. The dashed lines indicates the points on the surface that the observers see each spin period. These lines are perpendicular to $\mathbf{\Omega}$ and hence tilted by the angle $\hat{\theta}$ (artificially enhanced for clarity) with respect to the horizontal. The hotspot centres are indicated by the black dots.}
\label{figure10}
\end{figure*}

Figure \ref{figure10} shows a pulsar with two hotspots at (a) the beginning and (b) midway through a precession cycle. The hotspots are located at approximately $45^\circ, 225^\circ$ to match  Figures \ref{figure7} and \ref{figure9}. The shaded band indicates the region of the surface which is darker than average; the darkest points lie along the dot-dashed line. The band changes orientation from (a) to (b) as the star precesses rigidly about $\mathbf{e}_3$ (after a time $\pi/\dot{\psi}$ elapses).

To explain the flip in the fractional RMS at $\chi = 45^\circ$ (or equivalently $135^\circ$), we compare the observers at $\chi_1$ (north of $45^\circ$) and $\chi_2$ (south of $45^\circ$). In (a) and (b), $\chi_1$ traces similar paths close to the darkest and brightest regions respectively, leading to similar pulse amplitudes. However, the mean flux is lower in (a) than in (b), so the fractional RMS is a maximum at (a) and a minimum at (b). Similarly, $\chi_2$ traces a path with a greater pulse amplitude in (a) than in (b), but because these are normalised by the mean flux, which is greater at (a) than at (b), the fractional RMS is a minimum at (a) and a maximum at (b).

In order to explain the flip in mean flux between $65^\circ < \chi < 115^\circ$, we compare $\chi_2$ (south of $65^\circ$) and $\chi_3$ (south of $115^\circ$). In Figure \ref{figure10}(a), both $\chi_2$ and $\chi_3$ trace paths of similar brightness over one spin period. However in Figure \ref{figure10}(b), $\chi_2$ traces a path in the darker band, whereas $\chi_3$ traces a path closer to the bright centre of the hotspot. This confirms that the mean flux profiles seen by $\chi_2$ and $\chi_3$ are $\pi$ out of phase, as our simulations show.

\section{Triaxial star}
\label{triaxial}

A triaxial star is described by two separate ellipticity parameters, $\epsilon_1$ and $\epsilon_2$, defined by $\epsilon_1 = [2 (I_3 - I_1)/I_1]^{1/2}$ and $\epsilon_2 = [2 (I_3 - I_2)/I_2]^{1/2}$ \citep{zimsz79}. We pick $I_1$ to have a reference value of $I_0 = 0.4 M_{\star} R_{\star}^2$ and investigate the effect of varying $\epsilon_2/\epsilon_1$ from 0.3 to 0.7. Within this range, we find that there are no changes in the behaviour of $\Delta \Gamma_{\rm{phase-rms}}$, $\Delta \Gamma_{\rm{flux-rms}}$ or $B_{\rm{phase}}$. The only effect is to change the spin period, i.e. $P_{\rm{0.3}} \approx 1.0022 P_{\rm{0.5}}$, $P_{\rm{0.7}} \approx 0.9994 P_{\rm{0.5}}$ where $P_{\rm{0.3}}$ is the spin period for $\epsilon_2/\epsilon_1 = 0.3$ and so on.

The motion is discussed in detail by \citet{lanlif69}. When $J^2$ is only slightly larger than $2EI_1$, where $E$ is the total energy, the ${\mathbf{e}}_1$ axis rotates around $\mathbf{J}$ in an ellipse whose size increases as $J$ increases. As $J^2$ approaches $2EI_3$, ${\mathbf{e}}_3$ rotates around \textbf{J} in an ellipse. Hence $\theta$ is not constant: its mean, $\langle\theta\rangle$, increases with $J$. 

 We consider small $\theta$ in astrophysical problems, where $\theta$ now denotes the tilt angle at the start of the simulation. The angular velocity vector $\mathbf{\Omega}$ rotates periodically around the body axes, with period $T \sim 2 \pi/(\epsilon \Omega)$ (the exact value is given by a complete elliptic integral of the first kind). However, $\mathbf{\Omega}$ does not return to its original position with respect to the \textit{inertial} axes after one cycle. For a biaxial body, the spin frequency is simply $\dot{\phi}$. For a triaxial body, we have $\phi(t) = \phi_1(t) + \phi_2(t)$, where $\phi_1(t)$ has period $T$ and $\phi_2(t)$ has period $T'$ [incommensurable with $T$; \citep{lanlif69, zimsz79}]. 

We repeat the parameter searches in Sections \ref{single} and \ref{double} and find similar results for one and two hotspots. In some configurations, the precession phase profiles of $\Delta \Gamma_{\rm{phase-rms}}$, $\Delta \Gamma_{\rm{flux-rms}}$ and $B_{\rm{phase}}$ become non sinusoidal or contain strong harmonics. For consistency, we fit these profiles with the same format as before (see Section \ref{profiles}), recovering a modulation similar to Figures \ref{figure7}--\ref{figure9}.

The only noticeable difference is in the case of two hotspots. At $\chi = 90^\circ$, between $\alpha \approx \pm 30^\circ$ to $\pm 50^\circ$, $\Delta \Gamma_{\rm{flux-rms}}$ increases from 3.14 to 3.19.
There is also increased variation with $\phi(0)$ ($< 5\%$ for $\Delta \Gamma_{\rm{phase-rms}}$ and $< 7.5\%$ for $\Delta \Gamma_{\rm{flux-rms}}$).

\section{X-ray timing analysis}
\label{datareduction}
In Sections \ref{single}--\ref{triaxial}, we show that precession modulates the mean intensity and arrival times of pulses from one or two hotspots. In this section, we search for such modulation in X-ray timing observations of AMSPs made over recent years by the \textit{Rossi X-ray Timing Explorer} (\textit{RXTE}). We consider three sources observed by \textit{RXTE}: SAX J1808.4$-$3658, XTE J1814$-$338, and HETE J1900.1$-$2455. XTE 1900.1$-$2455 is unsuitable as it displays peculiar behaviour, including persistent X-ray emission and intermittent pulsations even during periods of low accretion \citep{gallowaymorg07}.  SAX J1808.4$-$3658 is also not ideal as the fractional RMS of the first harmonic changed erratically over the last four outbursts, and a previous analysis of the pulse-phase residuals did not reveal any periodicities consistent with precession \citep{hartman07}.
The most promising candidate is XTE J1814$-$338, whose data cover a 66-day outburst in which modulations in the mean flux, RMS, and pulse-phase residuals are visible by eye (see below).

\subsection{XTE J1814$-$338}
XTE J1814--338 is the fifth AMSP to be discovered, with a spin frequency of 314.4 Hz \citep{mank03}. Between MJD 52796 and 52834, the object experienced an ouburst during which 27 thermonuclear (type I) X-ray bursts were observed. This is the longest interval over which pulsations have been detected consistently. The data from this outburst were analysed previously, but with different emphases. \citet{watts05} and \citet{watts06} reported on the variability and energy dependence of these bursts, finding that the burst fractional amplitude (defined in the above papers) is constant during a burst and decreases with increasing photon energy. \citet{pap07} presented a timing analysis and refined orbital parameters previously published. They noted the modulation in the pulse-phase residuals and attributed it to movement of the accretion hotspot as the accretion rate varies.

\subsection{Observations}
\label{observations}
The source was observed by the \textit{RXTE} Proportional Counter Array [\textit{RXTE} PCA; \citet{jahoda96}], which consists of five proportional counter units (PCUs). During the course of an observation, different numbers of PCUs are turned on at different epochs, even within one data set. In order to accurately determine the background rate for each data set, the contribution from each PCU at all times must be tracked and accounted for. Data were collected in Event Mode over 64 energy channels (2--60 keV) with 125 $\mu$s time resolution.
The data comprise 91 internally contiguous blocks lasting from 2 to 30 ks and span a total of 66 days. The X-ray flux was measured by fitting a phenomenological model consisting of blackbody and power-law components, each attenuated by neutral absorption, to spectra extracted from Standard-2 mode data in the range 2.5-25 keV. The Standard-2 mode has 129 energy channels and 16\,s timing resolution.

\subsection{Timing analysis}

The data is processed using LHEASOFT\footnote{http://heasarc.gsfc.nasa.gov/lheasoft} version 5.3 (2003 November 17). We correct the photon times of arrival (TOAs) to the solar system barycentre. The X-ray flux is background subtracted using the \textit{RXTE}/PCA Mission-Long Bright Source background model\footnote{http://heasarc.gsfc.nasa.gov/docs/xte/pca\_news.html}. A separate response matrix is calculated for each observation to account for drift in the PCA gains. We remove all type I bursts from the data, subtracting all photons from 15 seconds before to 200 seconds after the burst peak; all the bursts have rise times of 1--8\,s and last from 100--200\,s \citep{watts05}. 

\begin{figure}
\includegraphics[scale=0.5]{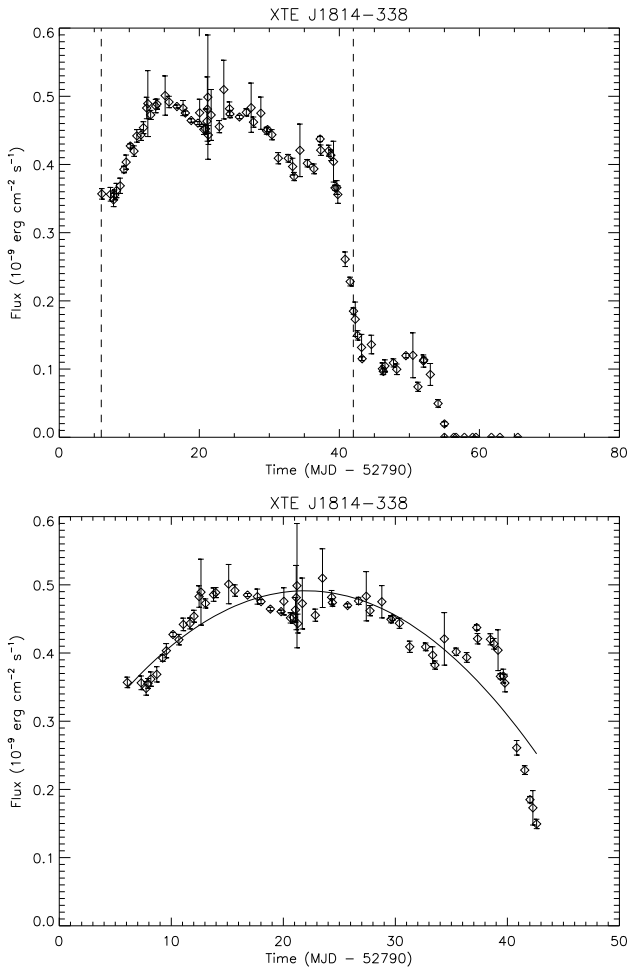}
\caption{Total flux (in units of 10$^{-9}$ erg/cm$^2$/s) versus time (Modified Julian Date). Top: The entire 66 day outburst starting at MJD 52796. Bottom: The first 37 days. The flux is measured with the \textit{RXTE} PCA in the energy band 2.5--25 keV. It is plotted with 1-$\sigma$ error bars. Overplotted is a quadratic polynomial which is later subtracted in order to extract the amplitude and period of the $\sim$ 12-day modulation (the `bumps' in the graph). }
\label{figure11}
\end{figure}

The X-ray flux is plotted versus time in Figure \ref{figure11}. It rises over the first five days from $3.5 \times 10^{-10}$ to $4.4 \times 10^{-10}$  erg\,cm$^{-2}$\,s$^{-1}$. It remains between $4.4 \times 10^{-10}$ and $5.1 \times 10^{-10}$ erg\,cm$^{-2}$\,s$^{-1}$ for the next 30 days, exhibiting strong modulations. It then drops sharply over the next two days and falls below the sensitivity threshold (0.2 $\times 10^{-10}$ erg\,cm$^{-2}$\,s$^{-1}$) on MJD 52834. We analyse the first 37 days of data only (between the dashed vertical lines in Figure \ref{figure11}).

The published orbital parameters at the time of writing \citep{markstank03, chung06} are incomplete: neither reference quoted the epoch of mean phase ($t_{90}$). To extract this quantity, we select a long barycentre-corrected data span (observation ID 80418-01-03-00, lasting 30.024 ks) and correct for the satellite orbit using a trial $t_{90}$ value. Any error in the trial $t_{90}$ value, or any other orbital parameter, can be extracted by comparing the actual and expected TOAs. The TOA residuals obey equation (3) in \citet{deeboy81}.

The fifth term on the right-hand side of the latter equation shows that the residuals in $t_{90}$ produce a sinusoidal variation in the TOAs, which we attempt to minimize. We also calculate corrections to the spin period by subtracting a linear trend from the TOAs, and the spin period derivatives by subtracting a quadratic trend [third term in \citet{deeboy81}]. We do not attempt to correct for other effects as these three terms dominate.
 The revised orbital parameters are quoted in Table \ref{tab:params}. 

\begin{table}
\begin{minipage}{70mm}
\renewcommand{\thefootnote}{\thempfootnote}
\caption{Orbital parameters for XTE J1814$-$338}
\begin{tabular}{|l|l|}
\hline
Barycentric spin period (s)& 0.003181105669954(4)\\
Spin frequency derivative (Hz\,s$^{-1}$)& $-7.2(3) \times 10^{-14}$\\
Projected semimajor axis (lt s)& 0.390626(2) \footnote{\citet{chung06}}\\
Epoch of $\pi/2$ mean phase (MJD) & 52808.8975258(4)\\
Orbital period (s) & 15388.7238(2) \footnotemark[\value{mpfootnote}]\\
\hline
\end{tabular}\label{tab:params}
\end{minipage}
\end{table}

\subsection{Pulse folding}

We fold the light curves for each contiguous observing segment at the measured spin period for that segment, taking into account the first-order spin frequency derivative.  

The folded pulse profiles are then fitted with a sinusoid comprising first and second harmonic components, as with the simulations above, viz. $a + b \sin (2 \pi \gamma + c) + d \sin (4 \pi \gamma + e)$, where $0 \leq \gamma \leq 1$ denotes the pulse-phase. Fitting is done using the Levenberg-Marquardt nonlinear least-squares algorithm. Again, from the fitted parameters, we measure the fractional RMS of the first and second harmonic ($b/\sqrt{2} a$ and $d/\sqrt{2} a$ respectively) as well as the pulse-phase residuals of the first and second harmonic ($\gamma_0 = 0.25 - c/2\pi$ and $\gamma_1 = 0.125 - d/4\pi$ respectively).  

Uncertainties in the fitted parameters are determined using the constant $\chi^2$ boundary method \citep{numrec}; i.e. $b, c, d$ and $e$ are iterated separately until $\chi^2$ increases by unity relative to its minimum. We find that the constant $\chi^2$ uncertainties are $\approx 1.5$ times the raw standard deviation from the least-squares fit, $\sigma_{\rm{fit}}$. Henceforth, to simplify the extensive analysis, we use $1.5 \sigma_{\rm{fit}}$ to qualify the uncertainties.

\subsection{First harmonic}
\label{firstharm}
 In order to clarify whether to include the second harmonic of the observed pulse profile in any further analysis, we take the fitted parameters of the first harmonic to be the `true' parameters and investigate the contribution of the second harmonic to the total signal. 
In the first 37 days of data, the first and second harmonic pulse-phase residuals $\gamma_0$ and $\gamma_1$ are similar. There is a $1.04\%$ difference in their gradients, and a linear trend between the two can be fitted with a slope of unity lying within the $4 \sigma$ limit. As for the fractional RMS, the first harmonic $b/\sqrt{2} a$ shows a $4.19 \times 10^{-4}$ day$^{-1}$ increase over the first 37 days, whereas the second harmonic $d/\sqrt{2} a$ decreases by $3.94 \times 10^{-4}$ day$^{-1}$. 

We calculate the Lomb-Scargle periodogram \citep{lomb76, numrec} for $\gamma_0$, $\gamma_1$, $b/\sqrt{2} a$ and $d/\sqrt{2} a$. The periodogram is discussed fully in Section \ref{timing}. For now, we merely note that a significant ($>4.5 \sigma$) 12.2-day periodic signal is found in $\gamma_0$ and $b/\sqrt{2} a$. In the second harmonic, this signal is present in $\gamma_1$ at a $3.5 \sigma$ level, but absent in $d/\sqrt{2} a$. In order to test the quality of the data, we fit 12.2-day sine waves to $b/\sqrt{2} a$ and $d/\sqrt{2} a$. The $3\sigma$ upper limit on the fractional amplitude of the fit in $d/\sqrt{2} a$ is 7.8\%, whereas the best fit fractional amplitude in $b/\sqrt{2} a$ is 6.9\%. We cannot therefore rule out the possibility that there is a hidden signal in the second harmonic fractional RMS.

For this reason, and since the first harmonic component of the fractional RMS dominates ($b/\sqrt{2} a \approx 0.103$, $d/\sqrt{2} a \approx 0.029$), we exclude the second harmonic fractional RMS from further consideration. We can also exclude $\gamma_1$ as it offers no additional information about the candidate precession signal. Nonetheless, in another object with a cleaner signal, or with better data, the pulse-phase residuals can yield extra information in principle, e.g. about the detailed form of the surface intensity map \citep{hartman07}.

\citet{hartman07} discussed how `red noise' (long-time-scale correlations) affect each harmonic's fractional RMS and pulse-phase residuals, causing them to vary independently. The authors use a common phase residual which is a weighted combination of $\gamma_0$ and $\gamma_1$, to minimise the intrinsic variation. We do not employ this technique as the same modulation appears in $\gamma_0$ and $\gamma_1$, and it is not detectable in the second harmonic fractional RMS.

\subsection{Searching for precession}
\label{timing}

\begin{figure}
\includegraphics[scale=0.55]{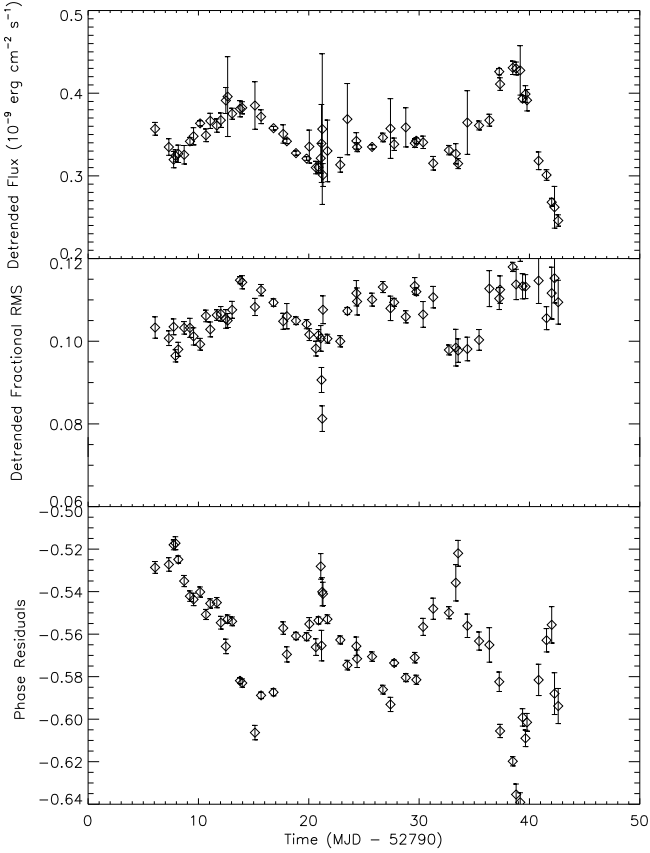}
\caption{Top: flux time series (folded on the spin period) after subtracting the long-term quadratic trend in Figure \ref{figure11}. Center:  fractional RMS of the first harmonic component of the folded pulse profile after folding on the spin period. Bottom: Pulse-phase residuals of first harmonic. All quantities are in the energy band 2.5--25 keV and are plotted with 1-$\sigma$ error bars, derived as explained in Section \ref{observations}.}
\label{figure12}
\end{figure}

\begin{figure}
\includegraphics[scale=0.55]{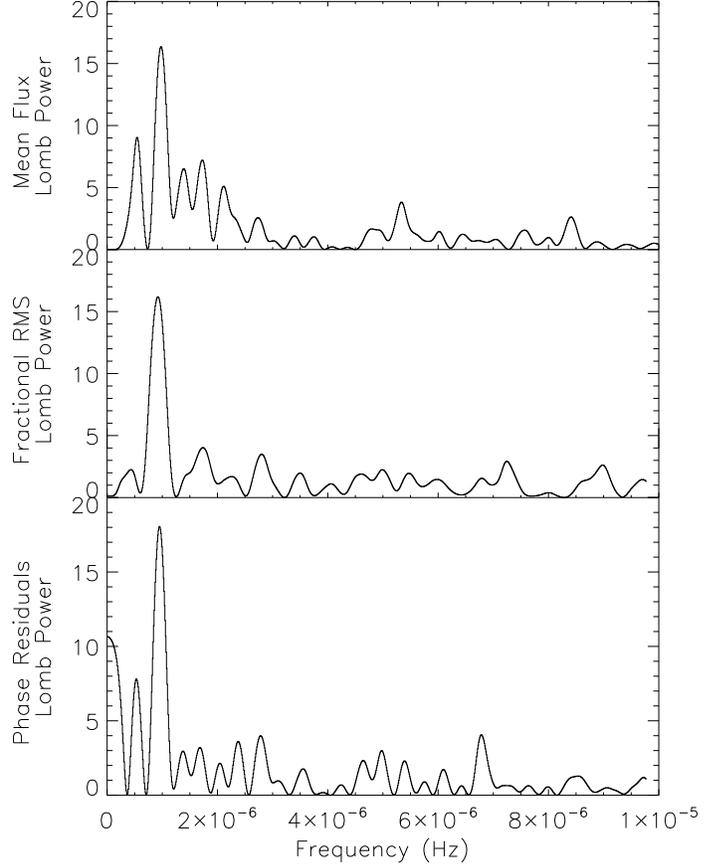}
\caption{Lomb periodogram for the following time series. Top: mean flux. Center: fractional RMS. Bottom: pulse-phase residuals. Frequencies are measured in Hz.}
\label{figure13}
\end{figure}

The three remaining quantities of interest are the total flux, fractional RMS and pulse-phase residuals of the first harmonic. As any longer period variation can mask shorter periods, a quadratic trend in the flux and a linear trend in the fractional RMS are subtracted, leaving the time series in Figure \ref{figure12}. To search for periodicities in these quantities, we construct a Lomb-Scargle periodogram, which calculates the significance of periodicities in unevenly sampled data \citep{lomb76, numrec}. The periodogram is plotted in Figure \ref{figure13}. The peak Lomb power is 18.1 for $\gamma_0$, 16.2 for $b/\sqrt{2}a$, and 16.4 for $a$. These Lomb powers correspond to a significance (the probability of a falsely detected signal) of order $10^{-6}$ for $\gamma_0$ and $10^{-5}$ for the other two quantities. The peak Lomb powers occur at periods $P_{\rm{flux}} = 283 \pm 22$ hours, $P_{\rm{phase}} = 293 \pm 21$ hours, and $P_{\rm{RMS}} = 302 \pm 23$ hours respectively (refer to Figure \ref{figure13}). The uncertainties in these periods are obtained in Monte-Carlo fashion by adding quasi-random noise (a Gaussian distribution with the same standard deviation as the original data minus a pure sine wave with the respective periods $P_{\rm{flux}}, P_{\rm{phase}}$ and $P_{\rm{RMS}}$) to noiseless signals with periods $P_{\rm{flux}}, P_{\rm{phase}}$ and $P_{\rm{RMS}}$.

\begin{table}
\caption{Results of fitting the folded mean flux, fractional RMS, and pulse-phase residuals (folding period = 292.59 hours) with the expression $A + B\sin(2\pi \Gamma + C)$, where $\Gamma$ is the precession phase. We list the parameters relevant to the precession model, i.e. the amplitude ($B_{\rm{rms}}$) and precession-phase offset ($C_{\rm{rms}}$) of the fractional RMS, the amplitude ($B_{\rm{phase}}$) and precession-phase offset ($C_{\rm{phase}}$) of the pulse-phase residuals, the precession-phase offset of the mean flux ($C_{\rm{flux}})$, the relative precession phase of the fractional RMS and pulse-phase residuals ($\Delta \Gamma_{\rm{phase-rms}}$), and relative precession phase of the fractional RMS and mean flux ($\Delta \Gamma_{\rm{flux-rms}}$).  Phases are expressed in radians.}
\begin{tabular}{|l|l|}
\hline
$B_{\rm{rms}}$ &  0.006 $\pm$ 0.003 \\
$B_{\rm{phase}}$ & 0.024 $\pm$ 0.003 \\
$C_{\rm{rms}}$ &  $-2.91 \pm$ 0.11 \\
$C_{\rm{phase}}$ & 0.12 $\pm$ 0.12  \\
$\Delta \Gamma_{\rm{phase-rms}}$ &3.1 $\pm$ 0.2 \\ 
$C_{\rm{flux}}$ & $-2.2 \pm$ 0.3  \\
$\Delta \Gamma_{\rm{flux-rms}}$ & 0.7 $\pm$ 0.3\\
\hline
\end{tabular}
\label{tab:rawvalues}
\end{table}

 The three periods are consistent and imply a mean candidate precession period of $P_p = 293 \pm 22$ hours.

\begin{figure}
\includegraphics[scale=0.55]{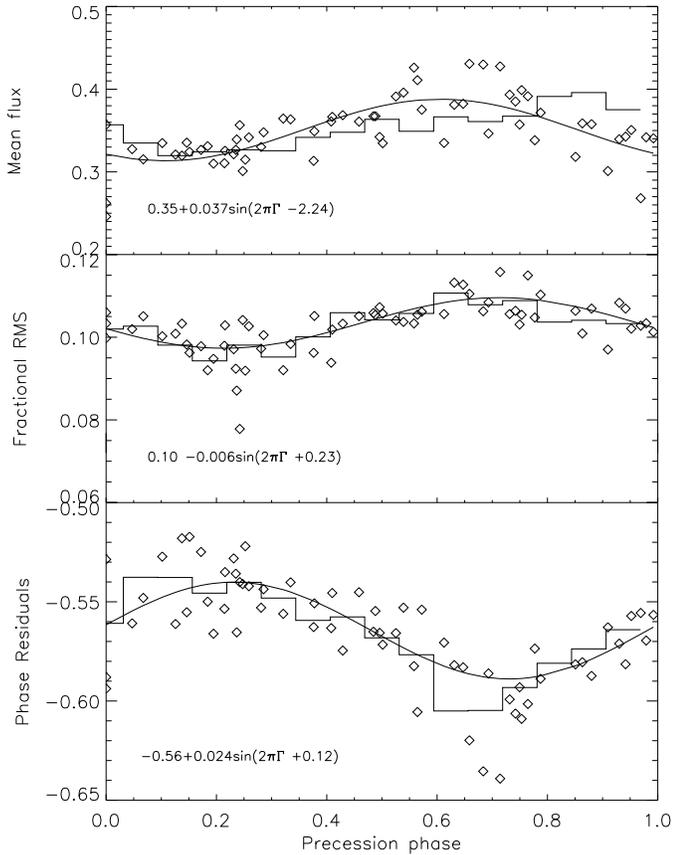}
\caption{Slow variation of the pulse characteristics \textit{refolded} on the candidate precession period $P_p = 293$ hr. Top: mean flux. Center: fractional RMS of the first harmonic. Bottom: pulse-phase residuals of the first harmonic. Measurements are represented by open symbols; the binned profile is represented by the histogram. The fitted sinusoid (with DC offset) is depicted as a solid curve.}
\label{figure14}
\end{figure}

We now fold the flux, fractional RMS, and pulse-phase residuals at $P_p$. The results are shown in Figure \ref{figure14}. We describe the resulting folded time series, as with the simulations, with a sinusoid $B \sin (2 \pi \Gamma + C)$ (where $0 \leq \Gamma \leq 1$ is the precession phase) plus a DC offset. The fitted parameters are listed in Table \ref{tab:rawvalues}. From these parameters, we can construct three quantities which are independent of mean flux and determined only by the orientation of the pulsar: the relative precession phase offset between the flux and the fractional RMS, $\Delta \Gamma_{\rm{flux-rms}}$, the relative precession phase offset between the pulse-phase residuals and the fractional RMS, $\Delta \Gamma_{\rm{phase-rms}}$; and the amplitude of the folded pulse-phase residual profile, $B_{\rm{phase}}$. The measured values are listed in Table \ref{tab:rawvalues}. Errors are obtained using the constant $\chi^2$ boundary method described above.

In addition to the quantities in Tables \ref{tab:rawvalues}, there is one arbitrary precession phase (corresponding to a choice of time origin) and two arbitrary amplitudes (for the mean flux and RMS), which depend on the unknown (and possibly changing) DC offset flux. The latter offset in general comes from orientation effects and also a DC component in the surface intensity map.

\section{Comparison between data and simulations}
\label{compare}
\begin{figure}
\includegraphics[scale=0.5]{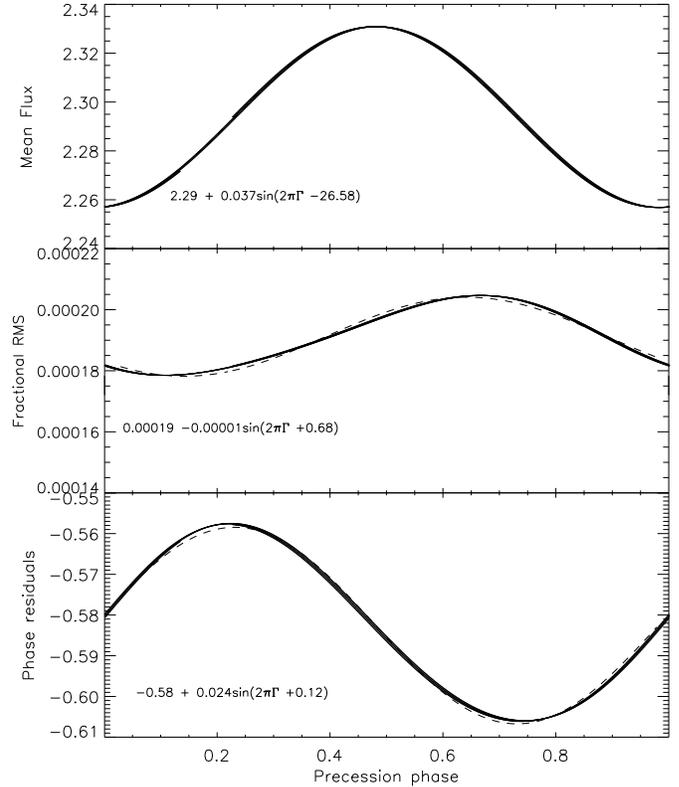}
\caption{Refolded simulated time series for the best match configuration ($\theta = 3^\circ, \chi = 179.95^\circ, \phi = 210^\circ, \alpha = 45^\circ$). $\Delta \Gamma_{\rm{flux-rms}} = 1.0$ and $B_{\rm{phase}} = 0.024$ match the data, but $\Delta \Gamma_{\rm{phase-rms}} = 2.6$ lies slightly outside the $1-\sigma$ error bar. Top: mean flux. Center: fractional RMS of the first harmonic. Bottom: pulse-phase residuals of the first harmonic. Simulated data are graphed as solid curves. Best-fit sinusoid is graphed as a dashed curve.}
\label{figure15}
\end{figure}

The only configuration for which the measured values of $\Delta \Gamma_{\rm{phase-rms}}, \Delta \Gamma_{\rm{flux-rms}}$ and $B_{\rm{phase}}$ come close to matching the data within experimental errors is a biaxial star with one hotspot. This agrees with the single-hemispheric-hotspot model suggested by \citet{muno02}.  For $\theta = 3^\circ, \chi = 179.95^\circ, \phi = 210^\circ, \alpha < 0^\circ$, we find $\Delta \Gamma_{\rm{phase-rms}}$ = 2.6, $\Delta \Gamma_{\rm{flux-rms}}$ = 1.0 and $B_{\rm{phase}}$ = 0.024. We plot $\Gamma_{\rm{flux}}, \Gamma_{\rm{rms}}$, and $\gamma_0$, refolded over the model's precession period, in Figure \ref{figure15}. The precession phase in the plots is offset to match the data (cf. Figure \ref{figure14}). The exact latitude of the hotspot does not affect these results, as mentioned previously (e.g. if $\alpha < 0^\circ$, then we would get the same results for $\chi = 0.05^\circ$). The above model for XTE J1814$-$338 is unlikely a priori. Given that pulsars are oriented randomly relative to an observer, the likelihood of observing a pulsar with $\chi < 1^\circ$ or $\chi > 179^\circ$ is 0.008\%. In fact, 95\% of the sky area covers the range $18^\circ \lesssim \chi \lesssim 162^\circ$. Secondly, for such a small inclination angle, $\mathbf{n}$ traces a small circle on the pulsar's surface during each rotation, resulting in a smaller ($<$ 1\%) fractional RMS than the $\sim$ 12\% level seen in the data.

For configurations with $1^\circ < \chi < 179^\circ$, our simulations fail to match the data. The relative precession phases of the fractional RMS, mean flux and phase residuals do not fall within $1\sigma$ of the measured values. Hence, if $\chi$ truly does lie in the above range, two scenarios are possible.

\subsubsection*{Scenario 1}
If the 293-hr modulation in the data is a real precession signal, our model is incomplete. For example, the intensity map may be more complicated in reality than equations (\ref{hotspot1}) or (\ref{hotspot2}, perhaps explaining the discrepancy between the simulations and the data in $\Delta \Gamma_{\rm{flux-rms}}$ and $\Delta \Gamma_{\rm{phase-rms}}$.

Unlike $\Delta \Gamma_{\rm{flux-rms}}$ and $\Delta \Gamma_{\rm{phase-rms}}$, it is possible to match $B_{\rm{phase}}$ for a relatively broad range of $\chi$. For one hotspot, the simulations match ($0.021 \leq B_{\rm{phase}} \leq 0.027$) for $6^\circ \leq \theta \leq 10^\circ$. For a triaxial star, the match occurs at certain combinations of $\theta$ and $\chi$ (see Figure \ref{figure7}), with the most probable combination drawn from $5^\circ \leq \theta \leq 8^\circ$ and $60^\circ \leq \chi \leq 120^\circ$. Although this agreement is insufficient as a proof of precession without an intensity map that also reproduces $\Delta \Gamma_{\rm{flux-rms}}$ and $\Delta \Gamma_{\rm{phase-rms}}$, it does provide some insight as to what the tilt angle would be in such a scenario.

Although we did not include the type I bursts in our analysis, the phase residuals of the burst oscillations are phase-locked with the non-accreting pulse phase residuals and are modulated on the same time-scale over the span of data that we use \citep{wattspat08}. This supports the precession model since precession of the entire pulsar would move the burst location(s) along with the non-accreting regions. The pulses from both areas would therefore be modulated in the same way.

Assuming a precession period of 293 hours and a tilt angle of $6^\circ$, equation (\ref{gw}) implies $\epsilon \sim 10^{-9}$ for XTE J1814--338. The gravitational wave strain $h_0$ at Earth from a biaxial rotator is given by 
\begin{equation}
h_0 = \frac{16 \pi^2 G}{c^4}\frac{\epsilon I_0 \Omega_{\ast}^2}{d},
\end{equation}
where $G$ is the gravitational constant, $c$ is the speed of light, $I_0$ is the star's moment of inertia and $d$ is the distance to the source. For XTE J1814$-$338, we have $d \approx 8 \pm 1.6$ kpc \citep{stroh03} and $Rc^2/GM > 4.2$ \citep{bhat05}, where $R$ is the star's radius, implying $10^{-28} \leq h_0 \leq 10^{-27}$.

\subsubsection*{Scenario 2}
If the 293-hr modulation is not due to precession [e.g. \citet{pap07} suggested that the hotspot drifts periodically around the star], then either the precession is heavily damped, making $\theta$ very small, or $\epsilon$ itself is smaller than expected.

If it is rigid, the star has a nonzero ellipticity for the reasons listed in Section \ref{intro}. However, in reality the star is elastic and probably contains a superfluid interior. Hence precession is damped via internal dissipation and gravitational radiation \citep{cutler01}. Internal dissipation generally dominates. Based on calculations by \citet{alpar88}, the time-scale for damping the tilt angle \citet{bondi55} is predicted to be between $400$ and $10^{4}$ precession periods. 
This effect or a small ellipticity, or some combination of both, lengthens the precession period beyond the $\sim37$-day observation window for $\epsilon \cos \theta < (P_\star/37$-d), i.e. $\epsilon \cos \theta \leq 9.9 \times 10^{-10}$. This implies $h_0 \leq 10 \times 10^{-27} \cos(\theta)^{-1}$.

\section{Conclusion}
\label{conclusion}
By analyzing X-ray timing data from the accreting millisecond pulsar XTE J1814$-$338, we find a 12.2-day periodicity in the mean flux, fractional RMS, and pulse-phase residuals of the first harmonic of the folded pulse. We measure two precession phase offsets relating these three quantities ($\Delta \Gamma_{\rm{phase-rms}} = 3.1 \pm 0.2$ rad and $\Delta \Gamma_{\rm{flux-rms}} = 0.7 \pm 0.3$ rad) as well as the amplitude of the pulse-phase residuals, $B_{\rm{phase}} = 0.024 \pm 0.003$.

Simulations of biaxial and triaxial precessing pulsars with one and two hotspots are also performed for a range of inclination angles ($0^\circ \leq \chi \leq 180^\circ$), tilt angles ($\theta \leq 10^\circ$), and hotspot latitudes ($-85^\circ \leq \alpha \leq 85^\circ$). 
$B_{\rm{phase}}$ is found to depend on the tilt angle, while $\Delta \Gamma_{\rm{phase-rms}}$ and $\Delta \Gamma_{\rm{flux-rms}}$ depend on the relative orientations of the line-of-sight and the hotspot(s). We find no significant dependence on the initial longitude at which the line-of-sight intersects the star except for small ($< 1^\circ$) inclination angles.

Comparing the data with the simulations, we are unable to find a model configuration  which matches the measured $\Delta \Gamma_{\rm{phase-rms}}$ or $\Delta \Gamma_{\rm{flux-rms}}$, unless we choose $0^\circ \leq \chi \leq 1^\circ$ (or $179^\circ \leq \chi \leq 180^\circ$), an a priori unlikely orientation. However, we are able to match $B_{\rm{phase}}$ for a range of tilt angles $5^\circ \leq \theta \leq 10^\circ$, if we are prepared to tolerate a discrepancy of $50^\circ$ in $\Delta \Gamma_{\rm{phase-rms}}$ and $55^\circ$ in $\Delta \Gamma_{\rm{flux-rms}}$ between the data and the model. One can therefore draw two possible conclusions: either the star is precessing but our surface intensity map is too simplistic, or the source is not precessing.
 If we attribute the 12.2-d periodicity to precession, this implies an ellipticity of $\epsilon \leq 3 \times 10^{-9}$, a gravitational wave strain $h_0 \leq 10^{-27}$, and hence a signal-to-noise ratio of $10^{-3}$ for initial LIGO and $10^{-2}$ for advanced LIGO (for a coherent 120-day search). On the other hand, if the precession is damped by internal dissipation ($\theta$ is small), or the precession period is much longer than the 37-day data span ($\epsilon$ is small), some other mechanism must cause the observed modulation. In this scenario, we find $\epsilon \cos \theta \leq 9.9 \times 10^{-10}$ and $h_0 \leq 10 \times 10^{-27} \cos(\theta)^{-1}$.

Although we face a negative result for this particular source, this paper establishes a framework for analyzing modulations in X-ray flux from AMSPs for a range of geometrical configurations and surface intensity maps. We anticipate that the framework will be applied to other AMSPs in the future. Given the values of $\epsilon$ inferred from the gravitational-wave stalling hypothesis \citep{bildsten98} and the theoretical models, e.g. of magnetic mountains \citep{pamel06, vigmel08}, it is clear that long-term X-ray monitoring of AMSPs (over years) is essential for predicting, and then searching for, their gravitational wave signal.

\section*{Acknowledgements}
We are grateful to Ranjan Singh for carrying out preliminary investigations of the precession model described in the manuscript. CC acknowledges the support of an Australian Postgraduate Award.

\bsp
\label{lastpage}
\end{document}